\documentclass[prl,aps,10pt,twocolumn,tightenlines,superscriptaddress, ]{revtex4-2}

\usepackage{bm,dcolumn,amsmath,graphicx,amsfonts,amssymb}
\usepackage{nicefrac}
\usepackage{color}
\usepackage{booktabs}
\usepackage{hyperref}
\usepackage[x11names]{xcolor}
\usepackage{widetable}
\usepackage{tablefootnote} 
\usepackage{adjustbox}
\usepackage{gensymb}
\definecolor{cite}{rgb}{0.,0.,0.9}
\hypersetup{colorlinks,linkcolor={cite},citecolor={cite},urlcolor={cite}}
\newcommand{\newcite}[1]{{\hypersetup{citecolor=red}\cite{#1}}}

\begin{document}

\title{
Interrogating the composition and distribution of nuclear magnetization via the hyperfine anomaly: experiment meets nuclear and atomic theory for short-lived \texorpdfstring{$^{47}$K}{47K}}

\author{M. L. Bissell} 
\email{mark.lloyd.bissell@cern.ch}
\affiliation{Experimental Physics Department, CERN, Geneva, Switzerland}

\author{M. Jankowski}
\affiliation{Experimental Physics Department, CERN, Geneva, Switzerland}
\affiliation{Technical University Darmstadt, Darmstadt, Germany}

\author{A. Antu{\v s}ek}
\affiliation{ATRI, Faculty of Materials Science and Technology in Trnava, Slovak University of Technology, Bratislava, Slovak Republic}

\author{N. Azaryan}
\affiliation{Experimental Physics Department, CERN, Geneva, Switzerland}

\author{B. C. Backes}
\affiliation{School of Physics, Engineering and Technology, University of York, York, United Kingdom}

\author{M. Baranowski}
\affiliation{Institute of Physics, Adam Mickiewicz University, Poznan, Poland}

\author{M. Chojnacki}
\affiliation{Experimental Physics Department, CERN, Geneva, Switzerland}
\affiliation{Nuclear and Particle Physics Department, University of Geneva, Geneva, Switzerland}

\author{K. M.  Dziubi{\'n}ska-K{\"u}hn}
\affiliation{Leipzig University, Leipzig, Germany}
\affiliation{Experimental Physics Department, CERN, Geneva, Switzerland}

\author{R. Han}
\altaffiliation{Present address: GSI, Darmstadt, Germany}
\affiliation{Department of Physics, University of Jyv{\"a}skyl{\"a}, Jyv{\"a}skyl{\"a}, Finland}

\author{A. Hurajt}
\affiliation{Department of Physical and Theoretical Chemistry, Faculty of Natural Sciences, Comenius University in Bratislava, Bratislava, Slovak Republic}
\affiliation{ATRI, Faculty of Materials Science and Technology in Trnava, Slovak University of Technology, Bratislava, Slovak Republic}

\author{B. Karg}
\affiliation{Nuclear and Particle Physics Department, University of Geneva, Geneva, Switzerland}



\author{I. Michelon}
\affiliation{Experimental Physics Department, CERN, Geneva, Switzerland}
\affiliation{Nuclear and Particle Physics Department, University of Geneva, Geneva, Switzerland}

\author{M. Pesek}
\affiliation{Experimental Physics Department, CERN, Geneva, Switzerland}

\author{M. Piersa-Si\l{}kowska}
\altaffiliation{Present address: Universidad Complutense de Madrid, Spain}
\affiliation{Experimental Physics Department, CERN, Geneva, Switzerland}

\author{B. M. Roberts}
\affiliation{School of Mathematics and Physics, The University of Queensland, Brisbane QLD, Australia}

\author{G. Sanamyan}
\altaffiliation{Present address: The School of Physics, Chemistry, and Earth Sciences, Adelaide University, Adelaide, Australia}
\affiliation{School of Mathematics and Physics, The University of Queensland, Brisbane QLD, Australia}

\author{T. P. Treczoks}
\affiliation{Department for Medical Physics and Acoustics, Carl von Ossietzky University Oldenburg, Oldenburg, Germany}

\author{L. V. Rodr\'iguez}  
\altaffiliation{Present address: Max-Planck-Institut f\"ur Kernphysik, Heidelberg, Germany.}
\affiliation{Experimental Physics Department, CERN, Geneva, Switzerland}

\author{H. Wibowo}
\affiliation{School of Physics, Engineering and Technology, University of York, York, United Kingdom}

\author{D. Zakoucky}
\affiliation{Czech Academy of Sciences, Rez, Czech Republic}

\author{M. {\v Z}{\v n}ava}
\affiliation{Department of Physical and Theoretical Chemistry, Faculty of Natural Sciences, Comenius University in Bratislava, Bratislava, Slovak Republic}

\author{M. Kortelainen}
\affiliation{Department of Physics, University of Jyv{\"a}skyl{\"a}, Jyv{\"a}skyl{\"a}, Finland}

\author{J. Dobaczewski}
\affiliation{School of Physics, Engineering and Technology, University of York, York, United Kingdom} 
\affiliation{Institute of Theoretical Physics, Faculty of Physics, University of Warsaw, Warsaw, Poland}

\author{J. S. M. Ginges}
\affiliation{School of Mathematics and Physics, The University of Queensland, Brisbane QLD, Australia}

\author{M. Kowalska} 
\affiliation{Experimental Physics Department, CERN, Geneva, Switzerland}
\affiliation{Nuclear and Particle Physics Department, University of Geneva, Geneva, Switzerland}

\date{\today}

\begin{abstract}

To date, the magnetic structure of nuclei has been poorly constrained, with limited information on its spatial distribution. In this work, we address the composition and distribution of nuclear magnetization in a precision study of short-lived $^{47}$K. 
We measure the Larmor frequency with part-per-million precision using liquid-state $\beta$-detected nuclear magnetic resonance at CERN-ISOLDE, improving determination of the experimental differential hyperfine anomaly relative to $^{39}$K by more than an order of magnitude.  
By combining these experimental results with relativistic all-orders atomic calculations and nuclear density functional theory, we obtain the relative spin and orbital contributions to the nuclear magnetic moments. Our analysis reveals an overestimation of the spin contribution predicted by nuclear theory, that persists even after considering two-body currents. Conversely, we show that the measured hyperfine anomaly is reproduced when adopting the spatial distribution of nuclear magnetization provided by density functional theory.  
The methodology introduced in this work establishes a means to probe the detailed magnetic structure of the nucleus. This is critical for benchmarking nuclear structure theory and calculations of symmetry-violating nuclear moments relevant to searches for physics beyond the Standard Model in atoms and molecules.

\end{abstract}
\maketitle

Precision measurements of nuclear electromagnetic properties—including spins, moments, and charge radii—form the foundation of our understanding of the nuclear force and many-body problem \cite{Cas90b,Ney03,nortershauser2023,Yang2023}. While these observables have successfully benchmarked sophisticated models \cite{Pastore2013,Reinhard2017,Bonnard2016}, fundamental discrepancies persist. Notably, the systematic reproduction of charge radii remains challenging \cite{garcia2016,Koszorus19,koszorus2021,perera2021,kortelainen2022,konig2023,karthein2024}, and to reconcile calculated magnetic moments with experiment, effective $g$ factors need to be introduced \cite{Richter2008}. Furthermore, the spatial distribution of neutrons \cite{grossman1999,horowitz2001, crystal_ball2014} and the evaluation of symmetry-violating nuclear moments -- such as the $CP$-violating Schiff and parity-violating anapole moments, which enter the interpretation of electric dipole moment and parity violation experiments in atoms and molecules, respectively  -- remain poorly constrained \cite{Ginges2004,Safronova2018,Arr24a,Ath25a}.

Addressing these challenges requires an alternative approach to the traditional nuclear probes. The Bohr-Weisskopf effect (BWE)~\cite{Bohr1950,Bohr1951} on the atomic hyperfine structure provides such a perspective by exploring the spatially extended magnetization distribution beyond the point-like limit. While many measurements of the BWE exist for stable nuclei~\cite{Persson2023} and the effect has recently been observed for the first time in a molecule \cite{Wilkins2025}, there is a notable absence of statistically significant data for radioactive isotopes with s or ms half-lives. Also, theoretical treatments have largely relied on simplified uniform distributions \cite{Dzuba1984} or phenomenological single-particle wave functions with adjusted parameters \cite{Demidov2023}.

In this work, we address the aforementioned shortcomings by introducing several new elements into the investigation of the BWE.
First, we measure the Larmor frequency of the short-lived \textcolor{black}{($t_{\frac{1}{2}}$=17.38 s)} nucleus $^{47}$K with part-per-million accuracy, using liquid-state $\beta$-detected Nuclear Magnetic Resonance (NMR) at the ISOLDE facility at CERN. This allows us to determine the precise values of the magnetic moment and the hyperfine anomaly (HA) relative to those of stable $^{39}$K.
Next, we perform state-of-the-art modeling of the nuclear magnetization distribution using a coherent microscopic approach, nuclear density functional theory (DFT)~\cite{Dob25g}, including, for the first time, the effects of two-body currents. Finally, we combine that approach with the most advanced determination of electronic correlations to establish the electron wave functions at the center of the atom, using the all-orders correlation potential method~\cite{DzubaCPM1989plaEn}. 

Thanks to the above ingredients, we can use the BWE, together with the magnetic moments, to constrain the composition and distribution of magnetization in the \textcolor{black}{$I^{\pi}=\frac{1}{2}^+$} $^{47}$K nucleus and the \textcolor{black}{$I^{\pi}=\frac{3}{2}^+$} reference isotope $^{39}$K. Because potassium has a single valence electron, it can be accurately addressed using atomic theory and efficiently polarized via optical pumping. Also, being one nucleon away from doubly-magic $^{40,48}$Ca, the magnetic properties of $^{39,47}$K can be addressed theoretically with high reliability \cite{Papuga2014}.

{\color{black}The hyperfine structure constant $\mathcal{A}$ quantifies the interaction strength of the atomic electron's non-uniform magnetic field with the nuclear magnetic moment.
The influence of the finite nuclear size on 
$\mathcal{A}$ is reflected in the HA, which -- in addition to the BWE \textcolor{black}{$\epsilon$} -- 
contains the Breit-Rosenthal \textcolor{black}{(BR)} effect {\color{black}$\delta$} due to the finite charge distribution.}
The total hyperfine constant is given by 
\begin{equation} 
\label{eqn:abs-anomaly}
\mathcal{A} = \mathcal{A}_0 (1+\epsilon) (1 +\delta) + \delta \mathcal{A}_{\rm QED}\,, 
\end{equation}
where $\mathcal{A}_0$ is the theoretical value for a point-like nucleus and $\delta \mathcal{A}_{\rm QED}$ quantum electrodynamic (QED) radiative corrections typically of a magnitude comparable to $\epsilon$.

Experimentally, the differential HA between isotopes $A$ and $A'$ can be determined with high precision from the ratio of their hyperfine constants and nuclear $g$ factors. \textcolor{black}{Further, for isotopes with similar charge radii and different spins, as is the case here, the differential BWE typically becomes the dominant contribution:}  
\begin{equation} 
\label{eqn:diff-anomaly}
^{A}\Delta^{A'} = \big(g^{(A')}/g^{(A)}\big) \big(\mathcal{A}^{(A)}/\mathcal{A}^{(A')}\big) - 1 \approx \epsilon^{(A)} - \epsilon^{(A')}. 
\end{equation} 
This observable effectively isolates variations in the magnetic distribution across an isotopic chain. Consequently, $^{A}\Delta^{A'}$ provides a stringent constraint for nuclear models, facilitating a direct test of theoretical descriptions of nuclear magnetization and its spatial evolution.

To derive the experimental differential HA $^{39}\Delta^{47}$, we determined a ratio of nuclear $g$ factors, $g(^{47}$K)/$g(^{39}$K), by comparing Larmor frequencies $\nu_L$ obtained via liquid-state $\beta$-NMR and conventional NMR, respectively. This ratio was then combined with the hyperfine constant $\mathcal{A}$ ratio derived from prior measurements \cite{Papuga2014}, allowing for the direct extraction of $^{39}\Delta^{47}$ through Eq.~(\ref{eqn:diff-anomaly}).

$^{47}$K was produced at the ISOLDE facility at CERN \cite{Borge2017} in a thick UC$_x$ target upon impact of 1.4 GeV protons. The atoms were subsequently surface ionized, accelerated to 50 keV and mass separated from other reaction products in the HRS separator.
After reaching the VITO beamline \cite{Kowalska2017, Croese2021,Jankowski2025}, the ion beam was neutralized via collisions with a vapor of potassium atoms in a charge exchange cell. The atomic beam of $^{47}$K was next polarized with $\sigma^+$ laser light, exciting the $4s\,{^2}$S$_{1/2}$ → $4p\,{^2}$P$_{3/2}$ transition (D2 line) at 766.5 nm. The polarized beam was then implanted into a liquid host, ionic liquid EMIM-DCA, located at the center of a 4.7 T magnet. 

\begin{figure}
    \centering
    \includegraphics[ trim={0.2cm 1.05cm 0.3cm 0.85cm},clip,width=1\columnwidth]{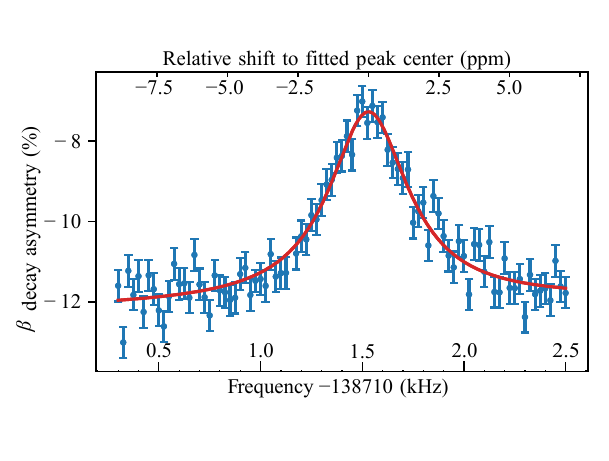}
    \caption{$\beta$-NMR spectrum of $^{47}$K in EMIM-DCA.}
    \label{fig:nmr}
\end{figure}

The $^{47}$K Larmor frequency was determined with high precision using liquid state $\beta$-NMR, as previously applied to the study of $^{26}$Na \cite{Harding2020,Jankowski2025}.  
The fast molecular tumbling in the liquid and sub-ppm field homogeneity allowed for extremely narrow $\beta$-NMR resonances with only a few ppm width, as shown in Fig. \ref{fig:nmr}.
Reference measurements were performed with in-situ conventional NMR on $^2$H nuclei in a heavy water sample, as described in \cite{Jankowski2025}.

From analysis of 20 $\beta$-NMR and NMR measurements, the frequency ratio $\nu_L(^{47}\mathrm{K,EMIM~DCA})$/ $\nu_L(^2\mathrm{H,D_2O}) =4.510701(2)$ was determined, which we then corrected for the different magnetic susceptibilities of our two samples (for details, see End Matter). \textcolor{black}{The uncertainty includes the statistical uncertainty and a systematic contribution due to small changes in the reference probe position and drift in the magnetic field between $^{47}$K and $^2$H reference measurements.}

To determine the $g$-factor ratio in Eq.~(\ref{eqn:diff-anomaly}) the above result may be combined with $g(^{39}$K) obtained from the conventional NMR measurement of $^{39}$K in the same environment as $^{47}$K. Unfortunately, due to a low NMR sensitivity of $^{39}$K, no clear resonance was visible in EMIM-DCA in our conventional-NMR measurements, even after 10 h of data taking.
Therefore, we used the literature ratio of $\nu_L(^{39}{\rm K})$ and $\nu_L(^{2}{\rm H)}$ in water, $0.30398485(9)$ \cite{Sahm1974}. Next, we corrected this value for the difference in the environments of $^{39}$K and $^{47}$K \cite{Sahm1974}, using state-of-the-art quantum chemistry calculations that provided the K NMR shielding of 1284(12)~ppm for water and 1247(10)~ppm for EMIM-DCA (for details, see End Matter).

\begin{figure}[h]
  \centering
  \includegraphics[trim={12cm 2.3cm 2cm 13cm},clip,
   width=\columnwidth] {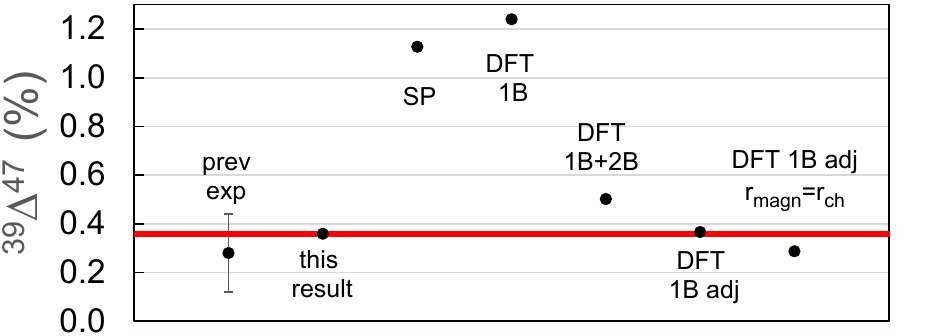}
  \caption{\label{fig:HFA-plot} Experimental and theoretical values of $^{39}\Delta^{47}$. Previous measurement \cite{Papuga2014} and our measurement. Theoretical values from Table \ref{tabP}, and DFT 1B adj, but with radial distribution of magnetization set to that of charge. The red band represents the experimental uncertainty.}
  \end{figure}

The corrected $\nu_L$ ratios lead to the $g$-factor ratio
$
g(^{47}\mathrm{K}) / g(^{39}\mathrm{K}) =   14.83802(23).
$
This value can be directly compared to the ratio of atomic hyperfine constants 
$\mathcal{A}(^{47}{\rm K})/\mathcal{A}(^{39}{\rm K})=14.785(1)${\color{black}\footnote{{\color{black}The uncertainty includes statistical contributions and a systematic contribution to $\mathcal{A}(^{47}{\rm K})$ from the asymmetric line shape.}}}
in the potassium $4s\,^2$S$_{1/2}$ ground state obtained via laser spectroscopy at ISOLDE and ABMR \cite{Papuga2014,Beckmann1974}.  Including both ratios in Eq.~(\ref{eqn:diff-anomaly}) produces the experimental differential HA between $^{47}$K and $^{39}$K, 
\begin{equation}
^{39}\Delta^{47}=0.35{\color{black}86}(1)(16)(68)\%.
\label{eqn:result}
\end{equation}
The first uncertainty arises from the Larmor-frequency determination, the second is due to uncertainties in NMR shielding related to the different chemical environments of EMIM-DCA and water, and the last due to uncertainty in $\mathcal{A}$ ratio. 
Given that the $4p\,^2$P$_{1/2}$ atomic level in potassium exhibits a negligibly small HA, our new value is in excellent agreement with that obtained via optical spectroscopy alone, $^{39}\Delta^{47}=0.28(16)\%$ \cite{Papuga2014}. However, as seen in Fig.~\ref{fig:HFA-plot}, our $\beta$-NMR measurement of $^{47}$K has allowed us to determine $\Delta$ with over 20-fold better precision. {\color{black}Further improvement would be possible with a more precise $\mathcal{A}$ ratio, as could be obtained from atomic rf-spectroscopy.}

We now turn to the theoretical values of the \textcolor{black}{BWE} ~\cite{Fujita1975,Shabaev1994}, expressed as $\epsilon= \epsilon_{\pi} + \epsilon_{\nu}$, with
\begin{equation}\begin{split}
\epsilon_{\pi} = 
-\frac{1}{\mu}\sum_{i=1}^{3}\Big[
b_{2i,S} &\langle g_S S_z r^{2i}\rangle_{\pi} + b_{2i,L} \langle  g_L L_z r^{2i}\rangle_{\pi}\\
&+(b_{2i,S}-b_{2i,L})\langle g_S Z_z r^{2i}\rangle_{\pi}
\Big] \
\label{BW}
\end{split}
\end{equation}
for the proton and similarly for the neutron. Here $S_z({\bf r})$ and $L_z({\bf r})$ refer to the $z$-projections of the spin and orbital nuclear densities. 
The expectation value is, e.g., $\langle g_S S_z r^{2j} \rangle_{\pi}=g_S^{\pi}\int dV {r}^{2j}S_{z,\pi}$, and the order of the radial moments of the magnetization distribution is specified by the value of $i$. We consider the dominating three terms in the sum over radial moments. The zero moment ($i=0$) corresponds to the spectroscopic nuclear magnetic dipole moment, $\mu = g_S^{\pi}\langle S_z\rangle_{\pi} + g_L^{\pi}\langle L_z\rangle_{\pi}+g_S^{\nu}\langle S_z\rangle_{\nu}$, and $g_S^{\pi,\nu}$ and $g_L^{\pi}$ are the free-nucleon spin and orbital $g$ factors. 
The term containing $Z_z=\sqrt{2\pi}[{\bf S}\times {\bf Y}^{(2)}]^{1}_0$ accounts for the spin asymmetry~\cite{Bohr1951}.   
The coefficients $b_{2i,S}$ and $b_{2i,L}$ are of electronic origin and {\color{black} describe the relative sensitivity of $\mathcal{A}$ to the radial moments of the nuclear spin and orbital magnetization distributions.
They arise from expansion of the terms \cite{Shabaev1994} } 
\begin{equation}\begin{split}
\int_0^{r}fg\, dR / \int_0^{\infty}fg\, dR&= \sum_{i=1} b_{2i,S}r^{2i}\, ,\\ 
\int_0^{r}fg(1-\frac{R^3}{r^3})\, dR / \int_0^{\infty}fg\, dR &= \sum_{i=1} b_{2i,L}r^{2i}\, ,\ 
\label{eq:expansion_ls}
\end{split}
\end{equation}
where, $f$ and $g$ are the upper and lower radial components of the atomic wave functions, normalized as $\int(f^2+g^2)R^2{\rm d}R=1$.
They are due to {\color{black}finite and non-uniform} electron wave functions {\color{black}within the nuclear volume and we find them numerically from a polynomial fit.
Atomic calculations were performed using the relativistic all-orders correlation potential method~\cite{DzubaCPM1989plaEn}; for more details, see End Matter.}

{\color{black} In the atomic calculations the wave functions are found in the Coulomb potential corresponding to a finite charge distribution (Fermi distribution with skin thickness $t = 2.3$ fm) and experimental $\langle r^2 \rangle$. In evaluating the hyperfine constants, the BR effect is thus accounted for, and we find $\delta(^{39}K)=0.2686\% \approx\delta(^{47}K)=0.2690\%$ for $4s\,^2S_{1/2}$, justifying the use of Eq.~\ref{eqn:diff-anomaly}.}


As seen in Eq.~(\ref{BW}), the BWE depends on the spin and orbital contributions to the magnetic moment and on their angular and radial distributions. Therefore, it provides additional information compared to the magnetic moment alone, which can be reproduced with an infinite number of combinations of spin and angular momentum contributions to the magnetic moment and is not sensitive to the distribution of magnetization.

The nuclear magnetization distributions and spin asymmetries were determined using the methodology recently developed in nuclear DFT \cite{Sas22c,Bon23c,Wib25d,Dob26}. 
We have used this approach to evaluate the symmetry-restored radial moments of the magnetization, with no effective $g$ factors. The results were obtained separately for the proton orbital, proton spin, neutron spin, and asymmetry contributions. Calculations were performed for $^{39}$K with an unpaired proton hole self-consistently blocked in the [202]3/2 deformed Nilsson orbital. For $^{47}$K, the relevant proton orbital is [200]1/2. 
Their intrinsic projections of angular momenta of $\Omega=+3/2$ and $\Omega=+1/2$, respectively, were aligned with the prolate-axial-symmetry axis. The spin and shape polarizations exerted by those holes generated self-consistent polarization of the calcium cores, with the total intrinsic magnetic dipole and electric quadrupole moments determined by the proton-hole and neutron- and proton-core contributions.
The spectroscopic moments of the $I=3/2^+$ and $I=1/2^+$ states, directly comparable to experimental data, were determined by restoring broken rotational symmetry {\color{black}\newcite{Rin80b,She21}}. 
{\color{black}In particular, without any effective charges used, the nuclear-DFT yields a fair description of the deformation properties of the potassium isotopes, with the calculated and measured \newcite{Pyykko2018} spectroscopic quadrupole moments of $^{39}$K, $Q_{\text{DFT}}=+0.0546$\,b v.s.\ $Q_{\text{exp}}=+0.0603(6)$\,b, and of $^{41}$K, $Q_{\text{DFT}}=+0.0607$\,b v.s.\ $Q_{\text{exp}}=+0.0734(7)$\,b.}

\begin{table}[t]
\caption{Values of $\mu$ and $\Delta$ for $^{39}$K and $^{47}$K. For theoretical values, the uncertainties in parentheses arise from varying the strength of the spin-spin interaction, with the Landau parameter ranging from 1.3 to 2.1 \cite{Sas22c}. Square brackets correspond to the RMS deviation among different DFT functionals.}
\begin{minipage}{\columnwidth}
\begin{ruledtabular}

\label{tabP}
\begin{tabular}{llll}
                     & $\mu(^{39}$K) ($\mu_N$)& $\mu(^{47}$K) ($\mu_N$)&  $^{39}\Delta^{47}$ ($\%$) \\  
 \hline
Experiment           &   +0.391464(5)       & +1.93618(4)          &  +0.359(7)                     \\
Single Particle\footnote{With rms magnetization radius equal to rms charge radius.}
                     &   +0.1243            & +2.7928              &   +1.127                       \\
DFT 1 body           &   +0.1259(2)[3]      & +2.66(3)[2]          &  +1.239(6)[47]                    \\
DFT 1B adj\footnote{ {\color{black}Italic values a}djusted to reproduce experimental magnetic moments of $^{39}$K and $^{47}$K.}
                     &   \textit{+0.391464}          & \textit{+1.93618}             &  +0.365[15]                         \\
DFT 1B +2B           &   +0.340[4]           & +2.87[2]              &  +0.502[22]                      \\

\end{tabular}
\end{ruledtabular}
\end{minipage}
\end{table}

Our experimental value and theoretical results, together with results of a single-particle model \cite{Bohr1950,Bohr1951}, are plotted in Fig. \ref{fig:HFA-plot} and shown together with magnetic-moment values in Table \ref{tabP}. One can see that our calculated magnetic moments for $^{39}$K and $^{47}$K remain close to the single-particle estimates but disagree substantially with the experimentally determined values.
{\color{black} Also}, the calculated differential HA is significantly larger than that measured and remains close to the single particle value. 

To investigate the origin of the above discrepancies between experimental and theoretical magnetic moments and differential HA, we first consider the strength of the spin and orbital contributions to both the $^{39}$K and $^{47}$K magnetic moments. In general, the experimental $\mu(^{39}$K) and $\mu(^{47}$K) alone do not provide any constraints on the composition of the magnetic moments, since they can be reproduced with any suitable combination of $g_S\langle S_z\rangle$ and $g_L\langle L_z\rangle$. 
In contrast, due to the different dependence of the BWE on these contributions, as seen in Eq. (\ref{eq:expansion_ls}), the experimental $^{39}\Delta^{47}$ and $\epsilon(^{39}K)$ (see End Matter) can provide a firm constraint on the composition of the magnetic moment.

\begin{figure}
   \centering
   \includegraphics[width=\columnwidth]{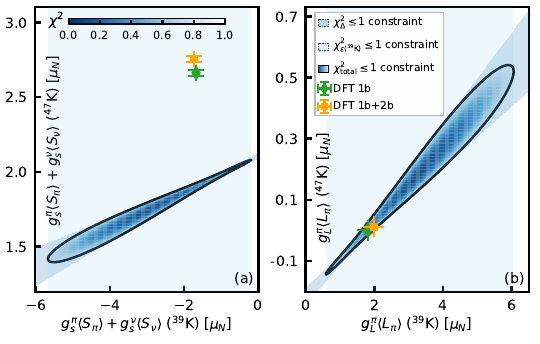}
   \vspace{-2mm}
   \caption{\label{fig:HFA-Rui-cut}Constraints on the spin (left) and orbital (right) angular momentum contributions to the magnetic moments of $^{39}$K and $^{47}$K based on the experimental values of their magnetic moments, the absolute HA of $^{39}$K, $\epsilon( ^{39}K)$, and the differential HA, $^{39}\Delta^{47}$. Theoretical values from DFT, including one- and two-body currents, are also shown.}
\end{figure}

The 1$\sigma$ confidence interval provided by these observables is represented by the black contour in Fig.~\ref{fig:HFA-Rui-cut}, in which
it can be seen that the calculated orbital contributions for both $^{39}$K and $^{47}$K agree within 1$\sigma$ of the experimental constraints, while the spin contributions do not present such an agreement. However, if we reduce the spin contributions of $^{39}$K and $^{47}$K so that theoretical magnetic moments agree with the experiment, we also reproduce our experimental differential HA within 1$\sigma$ (point `DFT 1 body adj' in Table \ref{tabP} and Fig.~\ref{fig:HFA-plot}). This suggests that here the spin contributions are responsible for the discrepancy between experiment and theory.

\begin{figure}
   \centering
   \includegraphics
   [trim={2.3cm 5.7cm 2.2cm 2.4cm},clip,
   width=\columnwidth]{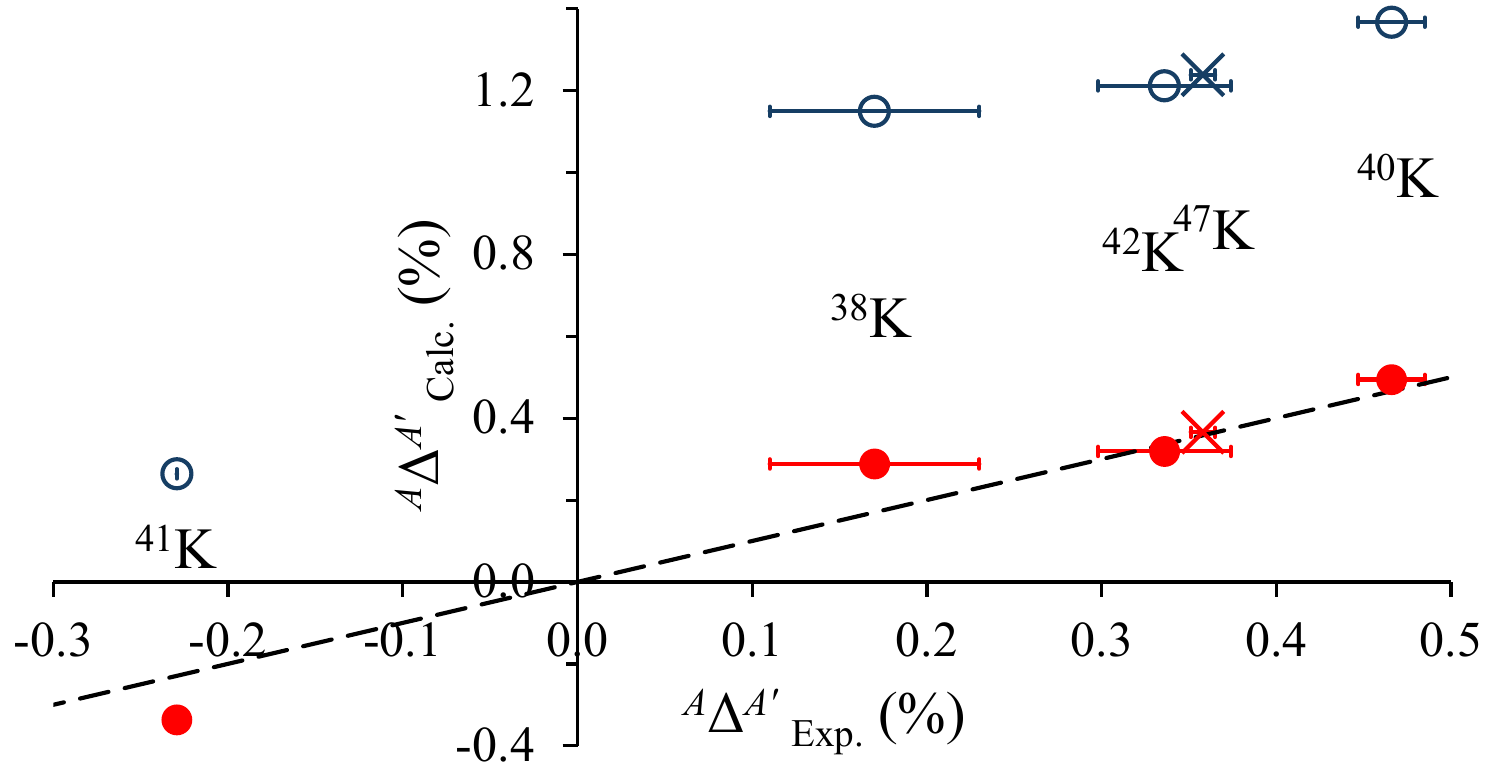}
   \caption{\label{fig:BW-HA-stable-isotopes}Differential HA between $^{39}$K and long lived $^{38-42}$K (circles), and the short-lived $^{47}$K (crosses) investigated in this work. Blue: direct DFT values. Red: values after scaling of only the spin contribution to reproduce the experimental magnetic moments. The dashed line shows where experiment and theory are in full agreement. }
   \end{figure}

To confirm the generality of this conclusion, we applied the same procedure to obtain the HA between $^{39}$K and other stable or long-lived potassium isotopes, for which the differential HA has been measured with sufficient precision. 
In Fig.~\ref{fig:BW-HA-stable-isotopes} we see that the calculated differential HA's show a much improved agreement with experiment after scaling the spin contributions.
Taken together, these observations strongly favor the hypothesis that the spin-like contributions are more generally responsible for both the discrepancies in the magnetic moments and the HA's. This observation is reminiscent of the empirical introduction of effective $g$ factors based on regional studies of magnetic moments~\cite{Richter2008}.

\begin{figure}
   \centering
    \includegraphics[width=\columnwidth]{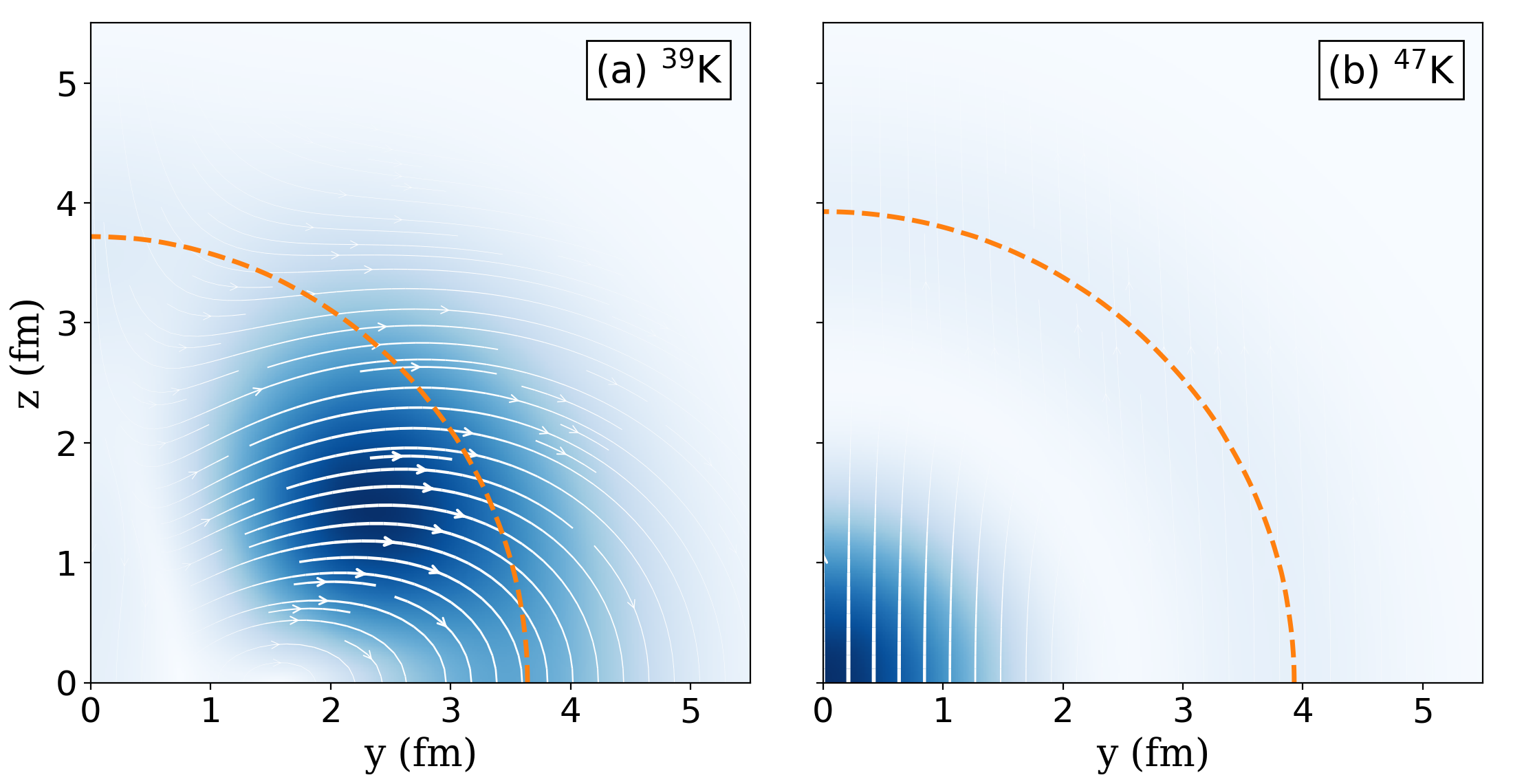}
    \caption{\textcolor{black}{\label{fig:muden}Calculated intrinsic magnetization distributions in $^{39}$K and $^{47}$K. Darker color indicates higher magnitude and arrows indicate direction. The dashed line indicates nuclear surface, i.e. matter density of $0.08\,{\rm fm}^{-3}$. Distributions are rotationally symmetric around the $z$-axis, and the scale is normalized separately for both nuclei. 
    }}
\end{figure}

It remains to examine the sensitivity of the differential HA to the radial distribution of magnetization. A frequently employed approximation when considering the BWE is the assumption that the magnetization distribution has a radial extent similar to the charge distribution \cite{Dzuba1984} and therefore $\langle r^2_{ S} \rangle=\langle r^2_{ L} \rangle=\langle r^2_{ch} \rangle$. Here, the calculated values $\langle r^2_{ S} \rangle=14.7$~fm$^2$ and$\langle r^2_{ L} \rangle=13.1$~fm$^2$ are significantly larger than the experimental mean squared charge radius of $^{39}$K $\langle r^2_{ch} \rangle=11.8$~fm$^2$. Consequently, the use of this approximation would have resulted in a predicted HA $^{39}\Delta^{47}=0.29\%$, which is some 10 sigma smaller than that observed, as seen in Fig.~\ref{fig:HFA-plot}.  
\textcolor{black}{To demonstrate that the calculated magnetization distributions are neither homogeneous, nor similar to charge distributions, we have plotted the intrinsic magnetization distributions in Fig. \ref{fig:muden}. The magnetization distribution in $^{39}$K shows a similarity to $0d_{3/2}$-orbit pattern, whereas the distribution in $^{47}$K has a spherical $1s_{1/2}$-orbit pattern. 
}   

To this point, all contributions to both the magnetic moments and HA's  were determined using the standard one-body magnetic-moment operator without effective $g$ factors. However, beginning with seminal early studies~\cite{Ma42, Vil47, Nym67, Nag71, Fujita1975}, contributions coming from the two-body meson exchange currents (MECs) have been proposed and evaluated by many authors. This question has recently been brought to new dimensions by the {\it ab initio} results obtained by Miyagi and collaborators~\cite{Miy24} who used the chiral effective field theory MEC magnetic dipole operator~\cite{Seu23}. Inspired by this study, Wibowo, Han, and collaborators~\cite{Wib26} have recently implemented calculations of the analogous contributions within nuclear DFT.
In this Letter, we determined the nuclear-DFT MEC contributions in $^{39}$K and $^{47}$K, 
presented in Fig. \ref{fig:HFA-plot} and Table \ref{tabP} (For details, see End Matter). The inclusion of two-body currents substantially improves the agreement between the experimental and theoretical magnetic moments of $^{39}$K, while worsening the agreement for $\mu(^{47}$K). The HA comes closer to the experimental value, but remains overestimated by 40\%. Thus, adding MECs in these two isotopes does not fully resolve the discrepancies between theory and experiment.

In summary, we have demonstrated that the HA, which is often neglected in nuclear structure studies, serves as a sensitive probe of the composition and spatial distribution of nuclear magnetic moments. We achieved this by combining high-resolution $\beta$-NMR spectroscopy with state-of-the-art atomic and nuclear theory, moving beyond historical approximations of the BWE.

Because the BWE exhibits a different sensitivity to the components of nuclear magnetization compared to the magnetic moments, we were able to independently isolate the spin and orbital contributions. Our analysis reveals that while the orbital magnetization is well-described, the spin component is significantly overestimated, a discrepancy that persists even with the inclusion of two-body currents. This finding provides a microscopic basis for addressing the long-standing necessity to use effective $g$ factors in reproducing experimental moments \cite{Richter2008}. Furthermore, we show that the experimental HA can be reproduced only with realistic spatial distributions {\color{black}and that we now have the experimental resolution to exclude simple models. Here we provide} the first validation that current DFT functionals accurately capture the spatial extent of nuclear magnetization.

Future investigations on short-lived isotopes near closed shells — where well-defined nuclear configurations minimize theoretical uncertainties ~\cite{Sas22c} — will be essential to refine this methodology. In particular, measurements of the differential BWE in $^{47-49}$K isotopes offer a promising avenue to probe predicted radially extended neutron distributions \cite{Bonnard2016}. 

More broadly, this work establishes a novel framework for mapping the (microscopic) composition and spatial extent of nuclear magnetization across the isotopic landscape, providing a robust tool to resolve long-standing challenges in nuclear structure. Furthermore, the ability to characterize the nuclear magnetization distribution offers a critical benchmark for nuclear structure models applied to calculations of symmetry-violating nuclear moments. Such moments are central to ongoing tests of the Standard Model and the search for physics beyond it.

\begin{acknowledgments}

\textit{Acknowledgments}--- We thank the ISOLDE Technical team for their support during beamtimes, the COLLAPS collaboration for the use of their tunable laser and optics, the University of York High Performance Computing service Viking and the Research Computing team for the computational support, the CSC-IT Center for Science Ltd., Finland, the IFT Computer Center at the University of Warsaw, Poland, and the Slovak NCC for HPC (311070AKF2) and HPC CLARA@UNIBA.SK of Comenius University, Bratislava, for providing computational resources. We also thank Dr. Marco Bonura from UNIGE for the EMIM-DCA bulk susceptibility measurement.
This work was supported by ERC Starting (No. 640645, BetaDropNMR) and Consolidator Grants (No. 101045813, PreSOBEN), CERN Experimental Physics Department, Wolfgang Gentner Programme of the German Federal Ministry of Education and Research (No. 13E18CHA), STFC Grants No. ST/W005832/1, ST/P003885/1, ST/W50791X/1, and ST/V001035/1, Leverhulme Trust Research Project Grant, Finnish Research Council (No. 339243), the Centre of Excellence in Neutron-Star Physics (No. 374066), Australian Research Council (ARC) Discovery Project No. DP230101685, the Ministry of Education, Youth and Sports of the Czech Republic (No. LM2023040), and during beamtimes by the ISOLDE Collaboration and the Horizon Europe research and innovation programme (No. 101057511).

\end{acknowledgments}
\bibliographystyle{apsrev4-2}
\bibliography{references}

@article{Richter2008,
  title = {$\mathit{sd}$-shell observables for the {USDA} and {USDB} Hamiltonians},
  author = {Richter, W. A. and Mkhize, S. and Brown, B. Alex},
  journal = {Phys. Rev. C},
  volume = {78},
  issue = {6},
  pages = {064302},
    year = {2008},
   doi = {10.1103/PhysRevC.78.064302},
  url = {https://link.aps.org/doi/10.1103/PhysRevC.78.064302}
}

@article{Reinhard2017,
  title = {Toward a global description of nuclear charge radii: Exploring the Fayans energy density functional},
  author = {Reinhard, P.-G. and Nazarewicz, W.},
  journal = {Phys. Rev. C},
  volume = {95},
  issue = {6},
  pages = {064328},
  numpages = {12},
  year = {2017},
  month = {Jun},
  doi = {10.1103/PhysRevC.95.064328},
  url = {https://link.aps.org/doi/10.1103/PhysRevC.95.064328}
}

@article{Bonnard2016,
  title = {Neutron Skins and Halo Orbits in the $sd$ and $pf$ Shells},
  author = {Bonnard, J. and Lenzi, S. M. and Zuker, A. P.},
  journal = {Phys. Rev. Lett.},
  volume = {116},
  issue = {21},
  pages = {212501},
  numpages = {5},
  year = {2016},
  month = {May},
  doi = {10.1103/PhysRevLett.116.212501},
  url = {https://link.aps.org/doi/10.1103/PhysRevLett.116.212501}
}

@article{Becker1082,
author = {Becker, D},
journal = {Z. Phys},
number = {37 a},
pages = {697},
title = {{Nuclear Magnetic Relaxation Induced by the Dynamics of Lattice}},
volume = {5},
year = {1982}
}

@article{Croese2021,
   author = {Croese, J and Baranowski, M and Bissell, M L and Dziubinska-K\"uhn, K M and Gins, W and Harding, R D and Jolivet, R B and Kanellakopoulos, A and Karg, B and Kulesz, K and others},
   doi = {https://doi.org/10.1016/j.nima.2021.165862},
   journal = {Nucl. Instrum. Meth. B},
   pages = {165862},
   title = {{High-accuracy liquid-sample beta-{NMR} setup at ISOLDE}},
   volume = {1020},
   url = {https://www.sciencedirect.com/science/article/pii/S0168900221008470},
   year = {2021}
}

@article{Ginges:2017fle,
  title = {Ground-state hyperfine splitting for {Rb, Cs, Fr, Ba$^+$, and Ra$^+$}},
  author = {Ginges, J. S. M. and Volotka, A. V. and Fritzsche, S.},
  year = {2017},
  journal = {Phys. Rev. A},
  volume = {96},
  number = {6},
  pages = {062502},
  doi = {10.1103/PhysRevA.96.062502},
  archiveprefix = {arXiv}
}

@article{Persson2010,
  title = {Hydrated metal ions in aqueous solution: How regular are their structures?},
  volume = {82},
  ISSN = {0033-4545},
  url = {http://dx.doi.org/10.1351/PAC-CON-09-10-22},
  DOI = {10.1351/pac-con-09-10-22},
  number = {10},
  journal = {Pure Appl. Chem.},
  author = {Persson,  Ingmar},
  year = {2010},
  month = aug,
  pages = {1901–1917}
}

@article{Demidov2023,
 title = {Bohr-Weisskopf effect in the potassium isotopes},
  author = {Demidov, Yu. A. and Kozlov, M. G. and Barzakh, A. E. and Yerokhin, V. A.},
  journal = {Phys. Rev. C},
  volume = {107},
  issue = {2},
  pages = {024307},
  numpages = {6},
  year = {2023},
  month = {Feb},
  publisher = {American Physical Society},
  doi = {10.1103/PhysRevC.107.024307},
  url = {https://link.aps.org/doi/10.1103/PhysRevC.107.024307}
}

@article{Jankowski2025,
doi = {10.1088/1748-0221/21/01/P01003},
url = {https://doi.org/10.1088/1748-0221/21/01/P01003},
year = {2026},
month = {1},
volume = {21},
pages = {P01003},
author = {Jankowski, M. and Azaryan, N. and Baranowski, M. and Bissell, M.L. and Brand, H. and Chojnacki, M. and Croese, J. and Dziubinska-K\"uhn, K.M. and Karg, B. and Madurga Flores, M. and others},
title = {Fully upgraded β-NMR setup at ISOLDE for high-precision high-field studies},
journal = {J. Inst.},
}

@article{Kowalska2017,
    title = {{New laser polarization line at the ISOLDE facility}},
    year = {2017},
    journal = {J. Phys. G},
    author = {Kowalska, M and Aschenbrenner, P and Baranowski, M and Bissell, M L and Gins, W and Harding, R D and Heylen, H and Neyens, G and Pallada, S and Severijns, N and others},
    number = {8},
    month = {8},
    pages = {084005},
    volume = {44},
    url = {http://dx.doi.org/10.1088/1361-6471/aa77d7 http://stacks.iop.org/0954-3899/44/i=8/a=084005?key=crossref.07759a8b6b4dbbf8d3c47fb20b5dae8e},
    doi = {10.1088/1361-6471/aa77d7},
    issn = {0954-3899}
}

@article{Borge2017,
  doi = {10.1088/1361-6471/aa5f03},
  url = {https://doi.org/10.1088/1361-6471/aa5f03},
  year = {2017},
  month = mar,
  volume = {44},
  number = {4},
  pages = {044011},
  author = {Maria J G Borge and Bj\"{o}rn Jonson},
  title = {{ISOLDE} past,  present and future},
  journal = {J. Phys. G}
}

@article{Sahm1974,
     author = {W Sahm and A Schwenk},
   journal = {Z. Naturforsch},
   pages = {1754-1762},
   title = {$^{39}${K}, $^{40}${K} and $^{41}${K} {Nuclear Magnetic Resonance Studies}},
   volume = {29},
   year = {1974}
}

@article{Papuga2014,
author = {Papuga, J. and Bissell, M. L. and Kreim, K. and Barbieri, C. and Blaum, K. and {De Rydt}, M. and Duguet, T. and {Garcia Ruiz}, R. F. and Heylen, H. and Kowalska, M. and others},
doi = {10.1103/PhysRevC.90.034321},
journal = {Phys. Rev. C},
number = {3},
pages = {034321},
title = {{Shell structure of potassium isotopes deduced from their magnetic moments}},
url = {https://link.aps.org/doi/10.1103/PhysRevC.90.034321},
volume = {90},
year = {2014}
}

@article{Harding2020,
author = {Harding, R. D. and Pallada, S. and Croese, J. and Antu{\v{s}}ek, A. and Baranowski, M. and Bissell, M. L. and Cerato, L. and Dziubinska-K{\"{u}}hn, K. M. and Gins, W. and Gustafsson, F. P. and others},
doi = {10.1103/PhysRevX.10.041061},
journal = {Phys. Rev. X},
number = {4},
pages = {041061},
title = {{Magnetic Moments of Short-Lived Nuclei with Part-per-Million Accuracy: Toward Novel Applications of $\beta$-Detected {NMR} in Physics, Chemistry, and Biology}},
url = {https://doi.org/10.1103/PhysRevX.10.041061 https://link.aps.org/doi/10.1103/PhysRevX.10.041061 http://arxiv.org/abs/2004.02820},
volume = {10},
year = {2020}
}

@article{Sas22c,
title = {Nuclear {DFT} analysis of electromagnetic moments in
         odd near doubly magic nuclei},
author = {P. L. Sassarini and J. Dobaczewski and J. Bonnard and R. F. {Garcia Ruiz}},
journal={J. Phys. G},
volume={49},
number={11},
pages={11LT01},
year = {2022},
url = {https://doi.org/10.1088/1361-6471/ac900a},
doi = {10.1088/1361-6471/ac900a}
}

@article{Bon23c,
title = {Nuclear {DFT} electromagnetic moments in heavy deformed open-shell odd nuclei},
journal = {Phys. Lett. B},
volume = {843},
pages = {138014},
year = {2023},
issn = {0370-2693},
doi = {10.1016/j.physletb.2023.138014},
url = {https://doi.org/10.1016/j.physletb.2023.138014},
author = {J. Bonnard and J. Dobaczewski and G. Danneaux and M. Kortelainen}
}

@article{She21,
        doi = {10.1088/1361-6471/ac288a},
        journal={J. Phys. G},
        volume={48},
        number={12},
        pages={123001},
        year = 2021,
        author = {Javid A Sheikh and Jacek Dobaczewski and Peter Ring and Luis Miguel Robledo and Constantine Yannouleas},
        title = {Symmetry restoration in mean-field approaches}
}

@article{Shabaev1994,
author = {Shabaev, V M},
journal = {J. Phys. B},
month = {dec},
pages = {5825},
volume = {27},
year = {1994},
url = {https://doi.org/10.1088/0953-4075/27/24/006},
}

@article{Fujita1975,
author = {Fujita, Takehisa and Arima, Akito},
journal = {Nucl. Phys. A},
pages = {513--541},
volume = {254},
year = {1975},
url = {https://doi.org/10.1016/0375-9474(75)90234-1},
}

@article{Bohr1951,
  title = {Nuclear Magnetic Moments and Atomic Hyperfine Structure},
  author = {Bohr, Aage},
  journal = {Phys. Rev.},
  volume = {81},
  issue = {3},
  pages = {331--335},
  numpages = {0},
  year = {1951},
  month = {Feb},
  doi = {10.1103/PhysRev.81.331},
  url = {https://link.aps.org/doi/10.1103/PhysRev.81.331}
}

@article{Bohr1950,
  title = {The Influence of Nuclear Structure on the Hyperfine Structure of Heavy Elements},
  author = {Bohr, Aage and Weisskopf, V. F.},
  journal = {Phys. Rev.},
  volume = {77},
  issue = {1},
  pages = {94--98},
  numpages = {0},
  year = {1950},
  month = {Jan},
  doi = {10.1103/PhysRev.77.94},
  url = {https://link.aps.org/doi/10.1103/PhysRev.77.94}
}

@article{Persson2023,
title = {Table of hyperfine anomaly in atomic systems - 2023},
journal = {Atom. Data Nucl. Data},
volume = {154},
pages = {101589},
year = {2023},
issn = {0092-640X},
doi = {https://doi.org/10.1016/j.adt.2023.101589},
url = {https://www.sciencedirect.com/science/article/pii/S0092640X23000177},
author = {J.R. Persson}
}

@article{Yang2023,
title = {Laser spectroscopy for the study of exotic nuclei},
journal = {Prog. Part. Nucl. Phys.},
volume = {129},
pages = {104005},
year = {2023},
issn = {0146-6410},
doi = {https://doi.org/10.1016/j.ppnp.2022.104005},
url = {https://www.sciencedirect.com/science/article/pii/S0146641022000631},
author = {X.F. Yang and S.J. Wang and S.G. Wilkins and R.F. Garcia Ruiz},
}

@article{Allegrini:2022xqva,
  title = {Survey of {{Hyperfine Structure Measurements}} in {{Alkali Atoms}}},
  author = {Allegrini, Maria and Arimondo, Ennio and Orozco, Luis A.},
  year = {2022},
  journal = {J. Phys. Chem. Ref. Data},
  volume = {51},
  number = {4},
  pages = {043102},
  doi = {10.1063/5.0098061}
}

@article{DzubaCPM1989plaEn,
  title = {Summation of the Perturbation Theory High Order Contributions to the Correlation Correction for the Energy Levels of the Caesium Atom},
  author = {Dzuba, V. A. and Flambaum, V. V. and Sushkov, O. P.},
  year = {1989},
  journal = {Phys. Lett. A},
  volume = {140},
  number = {9},
  pages = {493--497},
  doi = {10.1016/0375-9601(89)90129-1}
}

@article{Roberts:2022lda,
  title = {Electric Dipole Transition Amplitudes for Atoms and Ions with One Valence Electron},
  author = {Roberts, B. M. and Fairhall, C. J. and Ginges, J. S. M.},
  year = {2023},
  journal = {Phys. Rev. A},
  volume = {107},
  pages = {052812},
  doi = {PhysRevA.107.052812},
}

@article{Flambaum:2005ni,
  title = {Radiative Potential and Calculations of {{QED}} Radiative Corrections to Energy Levels and Electromagnetic Amplitudes in Many-Electron Atoms},
  author = {Flambaum, V. V. and Ginges, J. S. M.},
  year = {2005},
  journal = {Phys. Rev. A},
  volume = {72},
  number = {5},
  pages = {052115},
  doi = {10.1103/PhysRevA.72.052115}
}

@incollection{Cas90b,
  author={B. Castel and I. S. Towner},
  year={1990},
  title= {Modern theories of nuclear moments},
  publisher={Clarendon Press},
  editor={P. E. Hodgson},
  series={Oxford Studies in Nuclear Physics},
  volume={12},
  url={https://global.oup.com/academic/product/modern-theories-of-nuclear-moments-9780198517283}
}

@article{Ney03,
        doi = {10.1088/0034-4885/66/4/205},
        url = {https://doi.org/10.1088/0034-4885/66/4/205},
        year = {2003},
        month = {mar},
        volume = {66},
        number = {4},
        pages = {633--689},
        author = {Gerda Neyens},
        title = {Nuclear magnetic and quadrupole moments for nuclear structure research on exotic nuclei},
        journal = {Rep. Prog. Phys.}
}

@article{Arr24a,
doi = {10.1088/1361-6633/ad1e39},
url = {https://dx.doi.org/10.1088/1361-6633/ad1e39},
year = {2024},
volume = {87},
number = {8},
pages = {084301},
author = {Gordon Arrowsmith-Kron and Michail Athanasakis-Kaklamanakis
and Mia Au and Jochen Ballof and Robert Berger and Anastasia
Borschevsky and Alexander A Breier and Fritz Buchinger and Dmitry
Budker and Luke Caldwell and others},
title = {Opportunities for fundamental physics research with radioactive molecules},
journal = {Rep. Prog. Phys.}
}

@article{Ath25a,
  title = {Laser spectroscopy and {CP-violation} sensitivity of actinium monofluoride},
  author = {Athanasakis-Kaklamanakis, M. and Au, M. and Kyuberis, A. and Z\"ulch,
   C. and Gaul, K. and Wibowo, H. and Skripnikov, L. and Lalanne, L. and
   Reilly, J. R. and Koszor\'us, \'A and others},
  journal = {Nature},
  volume = {648},
  pages = {562-568},
  year = {2025},
  doi = {10.1038/s41586-025-09814-1},
  url = {https://doi.org/10.1038/s41586-025-09814-1}
}

@misc{Dob25g,
  title = {Electromagnetic and Exotic Moments in Nuclear {DFT}},
  author = {Dobaczewski,J. and Backes, B. C. and de Groote, R. P.  and Restrepo-Giraldo, A. and Sun, X.  and Wibowo, H.},
  year={2025},
  note={submitted to Annual Review of Nuclear and Particle Science},
  eprint={2511.04632},
  archivePrefix={arXiv},
  primaryClass={nucl-th},
  url={https://arxiv.org/abs/arXiv:2511.04632}
}

@article{Dob26,
  title = {Electromagnetic moments of ground and excited states calculated in heavy odd-$N$ open-shell nuclei},
  author = {Dobaczewski, J. and Stuchbery, A. E. and Danneaux, G. and Nagpal, A. and Sassarini, P. L. and Wibowo, H.},
  journal = {Phys. Rev. C},
  volume = {113},
  issue = {2},
  pages = {024306},
  numpages = {24},
  year = {2026},
  month = {Feb},
  doi = {10.1103/4q99-rv67},
  url = {https://link.aps.org/doi/10.1103/4q99-rv67}
}

@article{Wib25d,
doi = {10.1088/1361-6471/ade0dd},
url = {https://dx.doi.org/10.1088/1361-6471/ade0dd},
year = {2025},
month = {jul},
volume = {52},
number = {6},
pages = {065104},
author = {Wibowo, H and Backes, B C and Dobaczewski, J and de Groote, R P and Nagpal, A and S\'anchez-Fern\'ndez, A and Sun, X and Wood, J L},
title = {Electromagnetic moments in the {Sn-Gd} region determined within nuclear {DFT}},
journal = {J. Phys. G}
}

@article{Wib26,
  title = {The nuclear {DFT} two-body meson-exchange contributions to the magnetic dipole
  moments of atomic nuclei},
  author = {Herlik Wibowo and Rui Han and Bet\^ania C. Backes and Gauthier Danneaux and
Jacek Dobaczewski and Wick C. Haxton and Weiguang Jiang and Markus Kortelainen},
  journal = {},
  volume = {},
  pages = {},
  note = {to be published},
  year = {2026},
  doi = {},
  url = {}
}

@article{Dzuba1984,
doi = {10.1088/0022-3700/17/10/005},
url = {https://doi.org/10.1088/0022-3700/17/10/005},
year = {1984},
month = {may},
volume = {17},
number = {10},
pages = {1953},
author = {V A Dzuba and V V Flambaum and O P Sushkov},
title = {Relativistic many-body calculations of the hyperfine-structure intervals in caesium and francium atoms},
journal = {J. Phys. B - At. Mot. Opt.},
}

@article{Koszorus19,
  title = {Precision measurements of the charge radii of potassium isotopes},
  author = {Koszor{\'u}s, {\'A}. and Yang, X. F. and Billowes, J. and Binnersley, C. L. and Bissell, M. L. and Cocolios, T. E. and Farooq-Smith, G. J. and de Groote, R. P. and Flanagan, K. T. and Franchoo, S. and others},
  journal = {Phys. Rev. C},
  volume = {100},
  issue = {3},
  pages = {034304},
 year = {2019},
  doi = {10.1103/PhysRevC.100.034304},
  url = {https://link.aps.org/doi/10.1103/PhysRevC.100.034304}
}

@article{koszorus2021,
	title = {Charge radii of exotic potassium isotopes challenge nuclear theory and the magic character of {N=32}},
	volume = {17},
	issn = {1745-2481},
	url = {https://doi.org/10.1038/s41567-020-01136-5},
	doi = {10.1038/s41567-020-01136-5},
	number = {4},
	journal = {Nat. Phys.},
	author = {Koszor{\'u}s, {\'A}. and Yang, X. F. and Jiang, W. G. and Novario, S. J. and Bai, S. W. and Billowes, J. and Binnersley, C. L. and Bissell, M. L. and Cocolios, T. E. and Cooper, B. S. and others},
		year = {2021},
	pages = {439--443},
}

@article{kortelainen2022,
	title = {Universal trend of charge radii of even-even {Ca}--{Zn} nuclei},
	volume = {105},
	url = {https://link.aps.org/doi/10.1103/PhysRevC.105.L021303},
	doi = {10.1103/PhysRevC.105.L021303},
	number = {2},
	urldate = {2026-01-07},
	journal = {Phys. Rev. C},
	author = {Kortelainen, Markus and Sun, Zhonghao and Hagen, Gaute and Nazarewicz, Witold and Papenbrock, Thomas and Reinhard, Paul-Gerhard},
	month = feb,
	year = {2022},
	pages = {L021303},
}

@article{garcia2016,
	title = {Unexpectedly large charge radii of neutron-rich calcium isotopes},
	volume = {12},
	url = {https://www.nature.com/articles/nphys3645},
	doi = {10.1038/nphys3645},
		number = {6},
	journal = {Nat. Phys.},
	author = {Garcia Ruiz, R. F. and Bissell, M. L. and Blaum, K. and Ekström, A. and Frömmgen, N. and Hagen, G. and Hammen, M. and Hebeler, K. and Holt, J. D. and Jansen, G. R. and Kowalska, M. and others},
	year = {2016},
	pages = {594--598},
}

@article{pastore2013,
	title = {Quantum {Monte} {Carlo} calculations of electromagnetic moments and transitions in ${A \leq 9}$ nuclei with meson-exchange currents derived from chiral effective field theory},
	volume = {87},
	url = {https://link.aps.org/doi/10.1103/PhysRevC.87.035503},
	doi = {10.1103/PhysRevC.87.035503},
		number = {3},
		journal = {Phys. Rev. C},
	author = {Pastore, S. and Pieper, Steven C. and Schiavilla, R. and Wiringa, R. B.},
	month = mar,
	year = {2013},
	pages = {035503},
}

@article{perera2021,
	title = {Charge radii in covariant density functional theory: {A} global view},
	volume = {104},
	shorttitle = {Charge radii in covariant density functional theory},
	url = {https://link.aps.org/doi/10.1103/PhysRevC.104.064313},
	doi = {10.1103/PhysRevC.104.064313},
	number = {6},
	urldate = {2026-01-07},
	journal = {Phys. Rev. C},
	author = {Perera, U. C. and Afanasjev, A. V. and Ring, P.},
	month = dec,
	year = {2021},
	pages = {064313},
}

@incollection{nortershauser2023,
	title = {Nuclear {Charge} {Radii}},
	isbn = {978-981-15-8818-1},
	url = {https://link.springer.com/rwe/10.1007/978-981-15-8818-1_41-1},
	urldate = {2026-01-07},
	booktitle = {Handbook of {Nuclear} {Physics}},
	publisher = {Springer, Singapore},
	author = {Nörtershäuser, W. and Moore, I. D.},
	year = {2023},
	doi = {10.1007/978-981-15-8818-1_41-1},
	pages = {1--70},
}

@article{karthein2024,
	title = {Electromagnetic properties of indium isotopes illuminate the doubly magic character of {$^{100}$Sn}},
	volume = {20},
		url = {https://www.nature.com/articles/s41567-024-02612-y},
	doi = {10.1038/s41567-024-02612-y},
		number = {11},
		journal = {Nat. Phys.},
	author = {Karthein, J. and Ricketts, C. M. and Garcia Ruiz, R. F. and Billowes, J. and Binnersley, C. L. and Cocolios, T. E. and Dobaczewski, J. and Farooq-Smith, G. J. and Flanagan, K. T. and Georgiev, G. and {et. al}},
		year = {2024},
		pages = {1719--1725},
}

@article{konig2023,
	title = {Surprising Charge-Radius Kink in the {Sc} Isotopes at {N=20}},
	volume = {131},
	url = {https://link.aps.org/doi/10.1103/PhysRevLett.131.102501},
	doi = {10.1103/PhysRevLett.131.102501},
		number = {10},
		journal = {Phys. Rev. Lett.},
	author = {K{\"o}nig, Kristian and Fritzsche, Stephan and Hagen, Gaute and Holt, Jason D. and Klose, Andrew and Lantis, Jeremy and Liu, Yuan and Minamisono, Kei and Miyagi, Takayuki and Nazarewicz, Witold and others},
		year = {2023},
		pages = {102501},
}

@article{grossman1999,
	title = {Hyperfine {Anomaly} {Measurements} in {Francium} {Isotopes} and the {Radial} {Distribution} of {Neutrons}},
	volume = {83},
	url = {https://link.aps.org/doi/10.1103/PhysRevLett.83.935},
	doi = {10.1103/PhysRevLett.83.935},
		journal = {Phys. Rev. Lett.},
	author = {Grossman, J. S. and Orozco, L. A. and Pearson, M. R. and Simsarian, J. E. and Sprouse, G. D. and Zhao, W. Z.},
		year = {1999},
		pages = {935--938},
}

@article{horowitz2001,
	title = {Parity violating measurements of neutron densities},
	volume = {63},
	copyright = {http://link.aps.org/licenses/aps-default-license},
	issn = {0556-2813, 1089-490X},
	url = {https://link.aps.org/doi/10.1103/PhysRevC.63.025501},
	doi = {10.1103/PhysRevC.63.025501},
	number = {2},
	urldate = {2026-01-14},
	journal = {Phys. Rev. C},
	author = {Horowitz, C. J. and Pollock, S. J. and Souder, P. A. and Michaels, R.},
	month = jan,
	year = {2001},
	pages = {025501},
}

@article{crystal_ball2014,
	title = {Neutron {Skin} of $^{208}${Pb} from {Coherent} {Pion} {Photoproduction}},
	volume = {112},
	url = {https://link.aps.org/doi/10.1103/PhysRevLett.112.242502},
	doi = {10.1103/PhysRevLett.112.242502},
		number = {24},
		journal = {Phys. Rev. Lett.},
	author = {{Crystal Ball at MAMI and A2 Collaboration} and Tarbert, C. M. and Watts, D. P. and Glazier, D. I. and others},
	month = jun,
	year = {2014},
		pages = {242502},
}

@article{Schenck1996,
author = {Schenck, John F.},
title = {The role of magnetic susceptibility in magnetic resonance imaging: MRI magnetic compatibility of the first and second kinds},
journal = {Med. Phys.},
volume = {23},
number = {6},
pages = {815-850},
doi = {https://doi.org//10.1118/1.597854},
year = {1996}
}

@incollection{gupta_63_2007,
	address = {Berlin, Heidelberg},
	title = {63 {Diamagnetic} anisotropy of {H2O}},
	isbn = {978-3-540-23113-4 978-3-540-44694-1},
	url = {http://materials.springer.com/lb/docs/sm_lbs_978-3-540-44694-1_3044},
	urldate = {2026-01-20},
	booktitle = {Diamagnetic {Susceptibility} and {Anisotropy} of {Inorganic} and {Organometallic} {Compounds}},
	publisher = {Springer Berlin Heidelberg},
	author = {Jain, M. and Gupta, A.},
	editor = {Gupta, R. R.},
	year = {2007},
	doi = {https://doi.org/10.1007/978-3-540-44694-1_3044},
	pages = {3094--3094},
}

@article{Wilkins2025,
author = {S. G. Wilkins  and S. M. Udrescu  and M. Athanasakis-Kaklamanakis  and R. F. Garcia Ruiz  and M. Au  and I. Belošević  and R. Berger  and M. L. Bissell  and A. A. Breier  and A. J. Brinson  and others},
title = {Observation of the distribution of nuclear magnetization in a molecule},
journal = {Science},
volume = {390},
number = {6771},
pages = {386-389},
year = {2025},
doi = {10.1126/science.adm7717},
}

@article{Safronova2018,
  title = {Search for new physics with atoms and molecules},
  author = {Safronova, M. S. and Budker, D. and DeMille, D. and Kimball, Derek F. Jackson and Derevianko, A. and Clark, Charles W.},
  journal = {Rev. Mod. Phys.},
  volume = {90},
  issue = {2},
  pages = {025008},
  numpages = {106},
  year = {2018},
  month = {Jun},
  doi = {10.1103/RevModPhys.90.025008},
  url = {https://link.aps.org/doi/10.1103/RevModPhys.90.025008}
}

@article{SLy4,
	author = {E. Chabanat and P. Bonche and P. Haensel and J. Meyer and R. Schaeffer},
	doi = {https://doi.org/10.1016/S0375-9474(98)00180-8},
	issn = {0375-9474},
	journal = {Nucl. Phys. A},
	number = {1},
	pages = {231-256},
	title = {A Skyrme parametrization from subnuclear to neutron star densities Part II. Nuclei far from stabilities},
	url = {https://www.sciencedirect.com/science/article/pii/S0375947498001808},
	volume = {635},
	year = {1998}}

@article{SkXc,
	author = {Alex Brown, B.},
	doi = {10.1103/PhysRevC.58.220},
	issue = {1},
	journal = {Phys. Rev. C},
	month = {Jul},
	numpages = {0},
	pages = {220--231},
	title = {New Skyrme interaction for normal and exotic nuclei},
	url = {https://link.aps.org/doi/10.1103/PhysRevC.58.220},
	volume = {58},
	year = {1998}}

@article{SAMi,
	author = {Roca-Maza, X. and Col\`o, G. and Sagawa, H.},
	doi = {10.1103/PhysRevC.86.031306},
	issue = {3},
	journal = {Phys. Rev. C},
	month = {Sep},
	numpages = {6},
	pages = {031306},
	title = {New Skyrme interaction with improved spin-isospin properties},
	url = {https://link.aps.org/doi/10.1103/PhysRevC.86.031306},
	volume = {86},
	year = {2012}}

@article{UNEDF1,
	author = {Kortelainen, M. and McDonnell, J. and Nazarewicz, W. and Reinhard, P.-G. and Sarich, J. and Schunck, N. and Stoitsov, M. V. and Wild, S. M.},
	doi = {10.1103/PhysRevC.85.024304},
	issue = {2},
	journal = {Phys. Rev. C},
	month = {Feb},
	numpages = {15},
	pages = {024304},
	title = {Nuclear energy density optimization: Large deformations},
	url = {https://link.aps.org/doi/10.1103/PhysRevC.85.024304},
	volume = {85},
	year = {2012}}

@article{SIII,
	author = {M. Beiner and H. Flocard and Nguyen {Van Giai} and P. Quentin},
	doi = {https://doi.org/10.1016/0375-9474(75)90338-3},
	issn = {0375-9474},
	journal = {Nucl. Phys. A},
	number = {1},
	pages = {29-69},
	title = {Nuclear ground-state properties and self-consistent calculations with the skyrme interaction: (I). Spherical description},
	url = {https://www.sciencedirect.com/science/article/pii/0375947475903383},
	volume = {238},
	year = {1975}}

@article{SkOP,
	author = {Reinhard, P.-G. and Dean, D. J. and Nazarewicz, W. and Dobaczewski, J. and Maruhn, J. A. and Strayer, M. R.},
	doi = {10.1103/PhysRevC.60.014316},
	issue = {1},
	journal = {Phys. Rev. C},
	month = {Jun},
	numpages = {20},
	pages = {014316},
	title = {Shape coexistence and the effective nucleon-nucleon interaction},
	url = {https://link.aps.org/doi/10.1103/PhysRevC.60.014316},
	volume = {60},
	year = {1999}}

@article{Ma42,
  title = {Electromagnetic Properties of Nuclei in the Meson Theory},
  author = {Ma, S. T. and Yu, F. C.},
  journal = {Phys. Rev.},
  volume = {62},
  issue = {3-4},
  pages = {118-126},
  numpages = {0},
  year = {1942},
  month = {Aug},
  doi = {10.1103/PhysRev.62.118},
  url = {https://link.aps.org/doi/10.1103/PhysRev.62.118}
}

@article{Vil47,
  title = {On the Magnetic Exchange Moment for {${\mathrm{H}}^{3}$} and {${\mathrm{He}}^{3}$}},
  author = {Villars, Felix},
  journal = {Phys. Rev.},
  volume = {72},
  issue = {3},
  pages = {256-257},
  numpages = {0},
  year = {1947},
  month = {Aug},
  doi = {10.1103/PhysRev.72.256.2},
  url = {https://link.aps.org/doi/10.1103/PhysRev.72.256.2}
}

@article{Nym67,
title = {Magnetic moments and pion currents in nuclear matter},
journal = {Nucl. Phys. B},
volume = {1},
number = {9},
pages = {535-550},
year = {1967},
issn = {0550-3213},
doi = {https://doi.org/10.1016/0550-3213(67)90089-2},
url = {https://www.sciencedirect.com/science/article/pii/0550321367900892},
author = {E.M. Nyman}
}

@article{Nag71,
  title = {Evidence for Anomalous ${g}_{l}$ Factors of the Nucleons and the Mesonic-Exchange Effect},
  author = {Nagamiya, S. and Yamazaki, T.},
  journal = {Phys. Rev. C},
  volume = {4},
  issue = {5},
  pages = {1961-1964},
  numpages = {0},
  year = {1971},
  month = {Nov},
  doi = {10.1103/PhysRevC.4.1961},
  url = {https://link.aps.org/doi/10.1103/PhysRevC.4.1961}
}

@article{Miy24,
  title = {Impact of Two-Body Currents on Magnetic Dipole Moments of Nuclei},
  author = {Miyagi, T. and Cao, X. and Seutin, R. and Bacca, S. and Ruiz, R. F. Garcia and Hebeler, K. and Holt, J. D. and Schwenk, A.},
  journal = {Phys. Rev. Lett.},
  volume = {132},
  issue = {23},
  pages = {232503},
  numpages = {6},
  year = {2024},
  month = {Jun},
  doi = {10.1103/PhysRevLett.132.232503},
  url = {https://link.aps.org/doi/10.1103/PhysRevLett.132.232503}
}

@article{Seu23,
  title = {Magnetic dipole operator from chiral effective field theory for many-body expansion methods},
  author = {Seutin, R. and Hernandez, O. J. and Miyagi, T. and Bacca, S. and Hebeler, K. and K\"onig, S. and Schwenk, A.},
  journal = {Phys. Rev. C},
  volume = {108},
  issue = {5},
  pages = {054005},
  numpages = {17},
  year = {2023},
  month = {Nov},
  doi = {10.1103/PhysRevC.108.054005},
  url = {https://link.aps.org/doi/10.1103/PhysRevC.108.054005}
}

@article{Helgaker1998,
  title = {Ab Initio Methods for the Calculation of {NMR} Shielding and Indirect Spin-Spin Coupling Constants},
  volume = {99},
  ISSN = {1520-6890},
  url = {http://dx.doi.org/10.1021/cr960017t},
  DOI = {10.1021/cr960017t},
  number = {1},
  journal = {Chem. Rev.},
  author = {Helgaker,  Trygve and Jaszu\'nski,  Micha{\l}‚ and Ruud,  Kenneth},
  year = {1998},
  month = dec,
  pages = {293–352}
}

@article{Antusek2005,
  title = {Nuclear magnetic dipole moments from {NMR} spectra},
  volume = {411},
  ISSN = {0009-2614},
  url = {http://dx.doi.org/10.1016/j.cplett.2005.06.022},
  DOI = {10.1016/j.cplett.2005.06.022},
  number = {1–3},
  journal = {Chem. Phys. Lett.},
  author = {Antu{\v{s}}ek,  Andrej and Jackowski,  Karol and Jaszuński,  Micha{\l} and Makulski,  W{\l}odzimierz and Wilczek,  Marcin},
  year = {2005},
  month = aug,
  pages = {111–116}
}

@article{Mahler2011,
  doi = {10.1021/ic2018693},
  url = {https://doi.org/10.1021/ic2018693},
  year = {2011},
  month = dec,
  volume = {51},
  number = {1},
  pages = {425--438},
  author = {Johan M\"{a}hler and Ingmar Persson},
  title = {A Study of the Hydration of the Alkali Metal Ions in Aqueous Solution},
  journal = {Inorg. Chem.}
}

@article{Sengupta2021,
  title = {Parameterization of Monovalent Ions for the {OPC3,  OPC,  TIP3P-FB,  and TIP4P-FB} Water Models},
  volume = {61},
  ISSN = {1549-960X},
  url = {http://dx.doi.org/10.1021/acs.jcim.0c01390},
  DOI = {10.1021/acs.jcim.0c01390},
  number = {2},
  journal = {J. Chem. Inf. Model.},
  author = {Sengupta,  Arkajyoti and Li,  Zhen and Song,  Lin Frank and Li,  Pengfei and Merz,  Kenneth M.},
  year = {2021},
  month = feb,
  pages = {869–880}
}

@article{Glezakou2005,
  title = {Electronic structure,  statistical mechanical simulations,  and EXAFS spectroscopy of aqueous potassium},
  volume = {115},
  ISSN = {1432-2234},
  url = {http://dx.doi.org/10.1007/s00214-005-0054-4},
  DOI = {10.1007/s00214-005-0054-4},
  number = {2–3},
  journal = {Theor. Chem. Acc.},
  author = {Glezakou,  Vassiliki-Alexandra and Chen,  Yongsheng and Fulton,  John L. and Schenter,  Gregory K. and Dang,  Liem X.},
  year = {2005},
  month = dec,
  pages = {86–99}
}

@article{CanongiaLopes2012,
  doi = {10.1007/s00214-012-1129-7},
  url = {https://doi.org/10.1007/s00214-012-1129-7},
  year = {2012},
  volume = {131},
  number = {3},
  author = {Jose N. Canongia Lopes and Agilio A. H. Padua},
  journal = {Theor. Chem. Acc.},
}

@article{Eastman2017,
  doi = {10.1371/journal.pcbi.1005659},
  url = {https://doi.org/10.1371/journal.pcbi.1005659},
  year = {2017},
    volume = {13},
  number = {7},
  pages = {e1005659},
  author = {Peter Eastman and Jason Swails and John D. Chodera and Robert T. McGibbon and Yutong Zhao and Kyle A. Beauchamp and Lee-Ping Wang and Andrew C. Simmonett and Matthew P. Harrigan and Chaya D. Stern and others},
   title = {{OpenMM} 7: Rapid development of high performance algorithms for molecular dynamics},
  journal = {{PLOS} Comput. Biol.}
}

@inbook{IZAGUIRRE2009,
  title = {Multiscale dynamics of macromolecules using normal mode {Langevin}},
  ISBN = {9789814295291},
  url = {http://dx.doi.org/10.1142/9789814295291_0026},
  DOI = {10.1142/9789814295291_0026},
  booktitle = {Biocomputing 2010},
  publisher = {WORLD SCIENTIFIC},
  author = {Izaguirre,  J. A. and Sweet,  C. R. and Pande
,  V. S.},
  year = {2009},
  month = oct,
  pages = {240–251}
}

@article{Aaqvist2004,
  title = {Molecular dynamics simulations of water and biomolecules with a {Monte Carlo} constant pressure algorithm},
  volume = {384},
  ISSN = {0009-2614},
  url = {http://dx.doi.org/10.1016/j.cplett.2003.12.039},
  DOI = {10.1016/j.cplett.2003.12.039},
  number = {4–6},
  journal = {Chem. Phys. Lett.},
  author = {Åqvist,  Johan and Wennerstr\"{o}m,  Petra and Nervall,  Martin and Bjelic,  Sinisa and Brandsdal,  Bjørn O.},
  year = {2004},
  month = jan,
  pages = {288–294}
}

@article{PhysRevA.109.042815,
  title = {Nuclear magnetic dipole moments of $^{75}\mathrm{As},$ $^{121}\mathrm{Sb},$ and $^{123}\mathrm{Sb}$ from ab initio calculations of {NMR} shielding constants and existing {NMR} experiments},
  author = {Hurajt, Andrej and K\ifmmode \mbox{\c{e}}\else \c{e}\fi{}dziera, Dariusz and Kaczmarek-K\ifmmode \mbox{\c{e}}\else \c{e}\fi{}dziera, Anna and Antu\ifmmode \check{s}\else \v{s}\fi{}ek, Andrej},
  journal = {Phys. Rev. A},
  volume = {109},
  issue = {4},
  pages = {042815},
  numpages = {7},
  year = {2024},
  month = {Apr},
  doi = {10.1103/PhysRevA.109.042815},
  url = {https://link.aps.org/doi/10.1103/PhysRevA.109.042815}
}

@article{Jensen2014,
  doi = {10.1021/ct5009526},
  url = {https://doi.org/10.1021/ct5009526},
  year = {2014},
  month = dec,
  volume = {11},
  number = {1},
  pages = {132--138},
  author = {Frank Jensen},
  title = {Segmented Contracted Basis Sets Optimized for Nuclear Magnetic Shielding},
  journal = {J. Chem. Theory Comput.}
}

@article{Matthews2020,
  doi = {10.1063/5.0004837},
  url = {https://doi.org/10.1063/5.0004837},
  year = {2020},
  volume = {152},
  number = {21},
  author = {Devin A. Matthews and Lan Cheng and Michael E. Harding and Filippo Lipparini and Stella Stopkowicz and Thomas-C. Jagau and P{\'{e}}ter G. Szalay and J\"{u}rgen Gauss and John F. Stanton},
  title = {Coupled-cluster techniques for computational chemistry: {The CFOUR} program package},
  journal = {J. Chem. Phys.},
}

@article{Neese2020,
  doi = {10.1063/5.0004608},
  url = {https://doi.org/10.1063/5.0004608},
  year = {2020},
  month = jun,
  volume = {152},
  number = {22},
  author = {Frank Neese and Frank Wennmohs and Ute Becker and Christoph Riplinger},
  title = {The {ORCA} quantum chemistry program package},
  journal = {J. Chem. Phys.}
}

@article{Repisky2020,
  doi = {10.1063/5.0005094},
  url = {https://doi.org/10.1063/5.0005094},
  year = {2020},
  month = may,
  volume = {152},
  number = {18},
  author = {Michal Repisky and Stanislav Komorovsky and Marius Kadek and Lukas Konecny and Ulf Ekstr\"{o}m and Elena Malkin and Martin Kaupp and Kenneth Ruud and Olga L. Malkina and Vladimir G. Malkin},
  title = {{ReSpect}: Relativistic spectroscopy {DFT} program package},
  journal = {J. Chem. Phys.}
}

@article{Antusek2012,
  title = {Coupled cluster study of {NMR} shielding of alkali metal ions in water complexes and magnetic moments of alkali metal nuclei},
  volume = {532},
  ISSN = {0009-2614},
  url = {http://dx.doi.org/10.1016/j.cplett.2012.02.036},
  DOI = {10.1016/j.cplett.2012.02.036},
  journal = {Chem. Phys. Lett.},
  author = {Antu{\v{s}}ek,  Andrej and K{\c e}dziera,  Dariusz and Kaczmarek-K{\c e}dziera,  Anna and Jaszu{\'n}ski,  Micha{\l}},
  year = {2012},
  month = apr,
  pages = {1–8}
}

@book{multinuclear1987,
  doi = {10.1007/978-1-4613-1783-8},
  url = {https://doi.org/10.1007/978-1-4613-1783-8},
  year = {1987},
  publisher = {Springer {US}},
  editor = {Joan Mason},
  title = {Multinuclear {NMR}}
}

@article{Ginges2004,
title = {Violations of fundamental symmetries in atoms and tests of unification theories of elementary particles},
journal = {Phys. Rep.},
volume = {397},
number = {2},
pages = {63-154},
year = {2004},
issn = {0370-1573},
doi = {https://doi.org/10.1016/j.physrep.2004.03.005},
url = {https://www.sciencedirect.com/science/article/pii/S0370157304001322},
author = {J. S. M. Ginges and V. V. Flambaum}
}

@Article{Beckmann1974,
author={Beckmann, A.
and B{\"o}klen, K. D.
and Elke, D.},
title={Precision measurements of the nuclear magnetic dipole moments of6Li,7Li,23Na,39K and41K},
journal={Z. Phys.},
year={1974},
month={Sep},
day={01},
volume={270},
number={3},
pages={173-186},
issn={0044-3328},
doi={10.1007/BF01680407},
url={https://doi.org/10.1007/BF01680407}
}

@book{Rin80b,
  title={The nuclear many-body problem},
  author={Ring, Peter and Schuck, Peter},
  year={1980},
  publisher={Springer-Verlag, Berlin},
  url={https://www.springer.com/gp/book/9783540212065}
}

@article{Pyykko2018,
author = { Pyykk\"o, P},
title = {Year-2017 nuclear quadrupole moments},
journal = { Molecular Physics},
volume = {116},
number = {10},
pages = {1328--1338},
year = {2018},
publisher = {Taylor \& Francis},
doi = {10.1080/00268976.2018.1426131},
URL = { https://doi.org/10.1080/00268976.2018.1426131},
eprint = { {https://doi.org/10.1080/00268976.2018.1426131 }}

}

@article{Sanamyan2023,
  title = {Empirical Determination of the Bohr-Weisskopf Effect in Cesium and Improved Tests of Precision Atomic Theory in Searches for New Physics},
  author = {Sanamyan, G. and Roberts, B. M. and Ginges, J. S. M.},
  journal = {Phys. Rev. Lett.},
  volume = {130},
  issue = {5},
  pages = {053001},
  numpages = {6},
  year = {2023},
  month = {Feb},
  publisher = {American Physical Society},
  doi = {10.1103/PhysRevLett.130.053001},
  url = {https://link.aps.org/doi/10.1103/PhysRevLett.130.053001}
}

\pagebreak
\centerline{{\bf End Matter}}

\textit{NMR shielding}---To determine the $g$-factor ratio, NMR shielding of the potassium cation in EMIM-DCA ionic liquid (IL), $\sigma_{iso}$(IL), was modeled with computational chemistry methods~\cite{Helgaker1998,Antusek2005}. Unlike ions in aqueous solutions~\cite{Persson2010,Mahler2011,Glezakou2005}, there are no structural experimental data available for metallic cations in ionic liquids. Therefore, to determine the K$^+$ solvation structure in EMIM-DCA, we used classical force-field molecular dynamics (MD) with the TIP3P-FB force-field~\cite{Sengupta2021} for K$^+$ cation
and CL\&P force-field~\cite{CanongiaLopes2012} describing EMIM$^{+}$ and DCA$^-$ ions. A simulation box was prepared with periodic boundary conditions containing a K$^+$ cation, 400 EMIM$^+$ cations, and 401 DCA$^-$ anions. 
After initial energy minimization, the system was equilibrated in the NPT ensemble for 3~ns ($T=500$~K,  $P=1$~bar), followed by 7 ns ($T=298$~K). A 20~ns production run ($T=298$~K) was performed in the NVT ensemble, using the fixed average volume from the previous NPT equilibration. In the simulations, the temperature and pressure were controlled by Langevin thermostat~\cite{IZAGUIRRE2009} and Monte Carlo barostat~\cite{Aaqvist2004}, respectively. MD simulations were performed with the OpenMM package~\cite{Eastman2017}.  
The equilibrium structure of the first solvation shell around the K$^+$ cation in EMIM-DCA consists of 5 or 6 DCA$^-$ anions, each orienting their terminal nitrogen atoms towards the K$^+$. This corresponds to a sharp peak in the radial distribution function located at 2.75~\AA. In the second solvation shell, EMIM$^+$ cations are typically found directing their methyl tails towards the central K$^+$ cation. 
 
Following our previous approach~\cite{Harding2020,PhysRevA.109.042815}, NMR shielding constants of K$^{+}$ in K$^+$(DCA$^{-}$)$_n$ complexes were calculated using the non-relativistic coupled cluster method with non-iterative triple excitations [CCSD(T)] and Jensen's basis set series (pcSseg-$n$)~\cite{Jensen2014}. A tailored B3LYP(HFexch = 0.42) functional with re-scaled Hartree-Fock (HF) exchange admixtures was prepared. This functional, which reproduces non-relativistic CCSD(T) shieldings, was subsequently transferred to a four-component relativistic Dirac-Kohn-Sham (DKS) framework. The final NMR shielding constants were calculated using uncontracted Jensen's basis pcSseg-2 for potassium and pcSseg-1 for other atoms. All NMR shielding calculations were realized in the CFOUR~\cite{Matthews2020}, ORCA~\cite{Neese2020}, and ReSpect~\cite{Repisky2020} packages. 
The final $\sigma_{iso}(\rm IL)=1247(10)$ ppm was determined by averaging the DKS/B3LYP(HFexch = 0.42) shieldings for the first solvation shell structures extracted from 300 randomly selected snapshots along the MD simulation trajectory. 
$\sigma_{iso}(\rm H_2O)=1284(12)$ ppm was averaged from MD simulations with the TIP3P-FB force-field for water~\cite{Sengupta2021} using the same DKS/B3LYP(HFexch = 0.42) method. This value is consistent with the previous static calculations~\cite{Antusek2012}. The resulting chemical shift $\delta_{iso}$ = 37 ppm fits the observed span of potassium chemical shifts~\cite{multinuclear1987}. The error bars for $\sigma_{iso}$ values were obtained as a composite error~\cite{PhysRevA.109.042815} including the error of correlation and relativistic effects, non-additivity, basis-set incompleteness error, and statistical error from the structure ensemble. 

\begin{table}
 \caption{Nuclear expectation values that contribute to the BW effect, Eq.~(\ref{BW}), obtained with UNEDF1~\cite{UNEDF1}. Notation $\langle ...\rangle_{\pi+\nu}$ means $\langle ...\rangle_{\pi}+\langle ...\rangle_{\nu}$. Units: $\mu_N\,{\rm fm}^{2i}$.}
 \begin{ruledtabular}     
\begin{tabular}
{cc|cccccc}
             & $2i$                     &   $^{38}$K    &   $^{39}$K    &   $^{40}$K    &   $^{41}$K    &   $^{42}$K    &   $^{47}$K    \\
\hline 

$\langle g_s S_z r^{2i}    \rangle_{\pi+\nu}$     & 0
                         &   -0.542      &   -1.673      &   -2.149      &   -1.630      &   -0.416      &   2.663       \\
    & 2
                   &   -8.69       &   -24.64      &   -35.54      &   -23.81      &   -8.71       &   35.78       \\
   & 4
                 &   -179        &   -476        &   -751        &   -451        &   -228        &   731         \\
  & 6
                      &   -4834       &   -11813      &   -20019      &   -10859      &   -7253       &   17956       \\
                                            $\langle g_l L_z r^{2i}    \rangle_\pi$    & 0
                          &   1.807       &   1.799       &   0.844       &   1.788       &   -1.202     & 0.000 \\
                        &2&    23.11      &    23.56      &    11.04      &    23.16      &   -15.35      &    -0.28      \\
    &
                  4     &    400        &    409        &    189        &    393        &   -257        &    -7         \\
  &
                  6     &    9152       &    9214       &    4176       &    8488       &   -5476       &    -158       \\
$\langle g_s Z_z r^{2i}\rangle_{\pi+\nu}$    & 2
                      &    7.3        &    19.2       &    25.2       &    18.7       &   -4.2        &   -0.4        \\
   & 4 
                    &    140        &    359        &    523        &    341        &   -54         &   -10         \\
   &  6
                      &    3587       &    8730       &    13603      &    7967       &   -632        &    -305         \\
\end{tabular}
\label{table:Parameters}
\end{ruledtabular}
\end{table}

\textit{Magnetic susceptibility corrections}--The Larmor frequency ratios of $^{47}$K/$^2$H were corrected for the magnetic susceptibilities of the samples, which modify the external magnetic field at the nucleus by a factor $1/3(1-\alpha) \kappa$. Here $\kappa$ is the bulk volume susceptibility of the material and $\alpha$ is the shape factor that depends on the geometry of the sample and its orientation to the magnetic field \cite{Becker1082}.
$\kappa$(D$_2$O) = $-8.866$ ppm was obtained from literature \cite{gupta_63_2007}. That of non-degassed EMIM-DCA - which was assumed to change little under degassing - was determined by us with SQUID magnetometry, $\kappa$(EMIM-DCA)~=~$-6.33$ ppm. 
The geometrical factors $\alpha$ for our measurements at CERN were simulated with the CST Studio package, giving: $\alpha = 0.48$ for D$_2$O in the reference NMR tube (3.3 mm in diameter and 15 mm long tube perpendicular to $B_0$) and $\alpha =0.507$ for EMIM-DCA on a thin disk at $45\degree$    to $B_0$ \cite{Schenck1996}.

\textit{DFT calculations-}
Seven DFT functionals were used to determine the magnetic moment contributions and radial distributions: UNEDF1, SAMi, SLy4, SIII, SkM*, SkO’, SkXc \cite{UNEDF1,SAMi,SLy4,SIII,SkOP,SkXc}. In addition, their Landau parameter was varied between 1.3 and 2.1 \cite{Sas22c}. For two-body calculations, the spatial distribution of the MEC field depends quadratically on the nonlocal (exchange) spin densities and thus cannot be cast into the form of the one-body magnetization distributions used in Eq.~(\ref{BW}). Therefore, to determine the hyperfine anomaly $^{39}\Delta^{47}$ including MEC, we have tentatively attributed the $(L,S)=(0,1)$ component of the intrinsic part of the MEC contributions to the spin part and the $(L,S)=(1,0)$ component from two-body current of the Sachs term to the orbital contribution. Here, we assumed that their radial moments are identical to those of the one-body magnetization. These restrictions are the subject of a separate study.
Table \ref{table:Parameters} summarizes the calculated  spin and orbital contributions to the magnetic moments and radial distributions of magnetization of $^{38-42,47}$K using UNEDF1 functional.

\textit{Atomic structure calculations}--Atomic calculations are performed in the all-orders correlation potential method~\cite{DzubaCPM1989plaEn}. 
The starting approximation is relativistic Hartree-Fock (HF), and correlation corrections are accounted for through the addition of a correlation potential constructed using Feynman diagram techniques. This includes all-orders screening of the Coulomb interaction by core electrons and the hole-particle interaction. 
By adding the correlation potential, $\Sigma$, to the HF Hamiltonian, 
$(H_{\rm HF} + \Sigma)\psi_v = E_v\psi_v$, and solving, the correlation potential is also included to all orders. To gauge the impact of higher-order correlations, which we use to help quantify uncertainties, we also perform calculations with the correlation potential evaluated at just the second order of perturbation theory.

The hyperfine interaction leads to a modification of the wavefunctions of the core electrons, $\psi_c + \delta\psi_c$, leading to a correction to the HF potential, $\delta V$.
This gives the ``core polarization" correction to matrix elements, $\langle\psi_v|h_{\rm hfs} + \delta V|\psi_v\rangle$.
It is found in the first order in the hyperfine interaction by solving the set of hyperfine-perturbed HF equations self-consistently for all core states, 
$\left( H_{\rm HF} - E_c\right)\delta\psi_c = -\left( h_{\rm hfs} + \delta V - \delta E_c \right)\psi_c$.
We also account for structural radiation (hyperfine correction to the correlation potential) and the correction to normalization of the wavefunctions, which appear in the third order of perturbation theory.

We include Breit and radiative quantum electrodynamics (QED) effects in the calculations. 
The Breit correction corresponds to the magnetic and retardation correction to the electron-electron Coulomb interaction.
The QED correction to the hyperfine constants is estimated by rescaling results from Ref.~\cite{Ginges:2017fle}, while the QED corrections to energies are calculated using the radiative potential method~\cite{Flambaum:2005ni}.
The effects are small for K.

Finally, we estimate the contribution of missed correlation effects by introducing a scaling factor in front of the correlation potential, $\Sigma\to\lambda\Sigma$, which is tuned to reproduce experimental energies. 
Due to the already excellent agreement for the {\sl ab initio} energies, see Table~\ref{tab:energies}, these factors are very close to 1.
The difference between the scaled and unscaled calculations provides another handle on uncertainties. 
\textcolor{black}{For full details on the many-body calculations, we refer the reader to the review~\cite{Ginges2004} and to Refs.~\cite{Ginges:2017fle,Sanamyan2023} on implementation for the hyperfine structure and BWE; for estimating the uncertainties, see Ref.~\cite{Roberts:2022lda}.}

\begin{table}
\caption{Calculations of removal energies (cm$^{-1}$) for K, and comparison with experiment.}
\label{tab:energies}
\begin{ruledtabular}
\begin{tabular}{lrrrr}
          & $4s_{1/2}$  & $5s_{1/2}$  & $4p_{1/2}$  & $5p_{1/2}$  \\
    \hline
    HF    & 32370.5 & 13407.1 & 21006.5 & 10012.1 \\
    $\Sigma$    & 2565.4  & 554.6   & 1010.0  & 294.7   \\
    Breit & -0.8    & -0.3    & -2.4    & -0.8    \\
    QED  & -5.2    & -1.2    & 0.3     & 0.1     \\
    Total & 34929.8 & 13960.1 & 22014.3 & 10306.0 \\
    Expt.  & 35009.8 & 13983.3 & 22024.6 & 10308.4 \\
    $\Delta$ (\%)   & -0.2\%   & -0.2\%   & -0.05\%  & -0.02\% 
\end{tabular}
\end{ruledtabular}
\end{table}

The agreement for the $p_{1/2}$ states, where the BW effect is very small, is exceptionally good.
By comparing the point-like theory value (subtotal in Table~\ref{tab:A}) with the experimental hyperfine constant in $4s$, we may directly extract the BW effect, finding $\epsilon = +0.59(62)\%$,
where the uncertainty is from atomic theory.
The uncertainty in the final values is dominated by uncertainty in the structure radiation correction, estimated from the spread in its values across approximations.
For the $s$-states of K, this correction is larger and more unstable than typical, leading to a relatively large uncertainty of $0.6\%$.

\begin{table}
\caption{Calculated hyperfine constants (MHz) for $^{39}$K with point-like magnetization (i.e., BW effect is excluded). 
}
\label{tab:A}
\begin{ruledtabular}
\begin{tabular}{lrrrr}
             & \multicolumn{1}{r}{$4s_{1/2}$}  & \multicolumn{1}{r}{$5s_{1/2}$}  & \multicolumn{1}{r}{$4p_{1/2}$}  & \multicolumn{1}{r}{$5p_{1/2}$}  \\
    \hline
    HF       & 146.904        & 38.876    & 16.617     & 5.735     \\
    $\Sigma$       & 48.497         & 6.963     & 6.262      & 1.591     \\
    $\delta V$      & 34.737         & 9.074     & 4.888      & 1.651     \\
    Breit    & 0.232          & 0.053     & -0.004     & -0.001    \\
    QED~\cite{Ginges:2017fle}    & -0.306         & -0.073    & -0.001     &  0.000         \\
    SR+N       & -2.075         & -0.309    & -0.013     & 0.006     \\
    $(\lambda-1)\Sigma$   & 1.519          & 0.275     & 0.068      & 0.013     \\
    Subtotal & 229.5(14)      & 54.86(31) & 27.817(75) & 8.995(24) \\
    
    Expt.~\cite{Allegrini:2022xqva}     & 230.8598601(7) & 55.5(6)   & 27.793(71) & 9.01(17)  \\
\end{tabular}
\end{ruledtabular}
\end{table}
\begin{figure}[h]
    \centering
        \includegraphics[trim={0.25cm 0.25cm 0.25cm 0.25cm},clip,width=0.71\columnwidth]{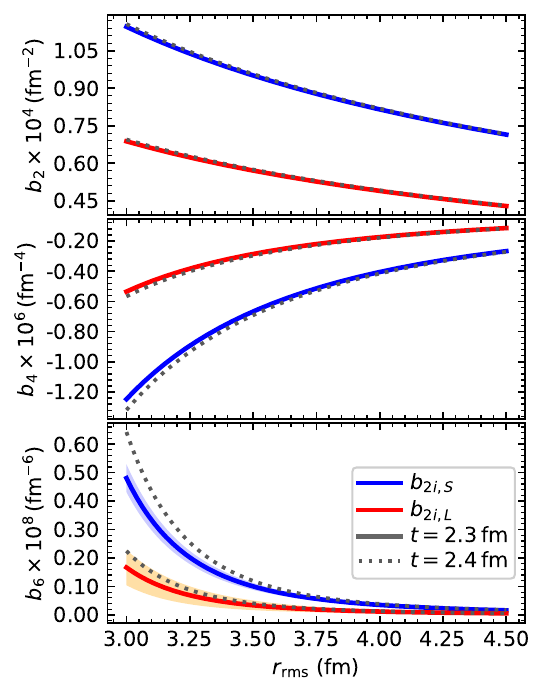}
    \caption{Atomic parameters $b_{2i}$ as a function of charge radii. 
    Shaded region shows numerical uncertainty, solid lines assume Fermi model skin thickness (90--10\% fall-off) $t=2.3\,{\rm fm}$; dashed line ($t=2.4\,{\rm fm}$) shows sensitivity.}
    \label{fig:b-values}
\end{figure}

The expansion coefficients of the electronic terms from Eq.~\eqref{eq:expansion_ls} for $^{38-42,47}$K are presented in Fig.~\ref{fig:b-values}.


\end{document}